\documentclass[aps,pra,preprint,11pt]{revtex4-1}
\usepackage{bm}
\usepackage{graphicx}
\usepackage[breaklinks=true,colorlinks=true,linkcolor=blue,urlcolor=blue,citecolor=blue]{hyperref}
\usepackage{subcaption}
\usepackage{bbm}
\usepackage{xcolor}
\expandafter\let\csname equation*\endcsname\relax
\expandafter\let\csname endequation*\endcsname\relax
\usepackage{amsmath,amsfonts,amssymb,amsthm}
\usepackage{braket}
\usepackage{mathrsfs}
\usepackage{mathtools}
\usepackage{array}
\newtheorem{theorem}{Theorem}[section]
\definecolor{myblue}{RGB}{26, 25, 140}
\definecolor{myred}{RGB}{221, 80, 109}

\newcommand{{\lambdaB}}{\bm\lambda}

\newcommand{\denM}[0]{\varrho}
\newcommand{\Tr}[1]{\text{Tr}\left\{#1\right\}}
\newcommand{\tr}[1]{\text{tr}\left\{#1\right\}}
\newcommand{\dd}[0]{\Delta}
\newcommand{\denp}[0]{\denM_{\bmp}}
\newcommand{\bmp}[0]{\bm{\lambda}}
\newcommand{\SLDp}[1]{L_{#1}}
\newcommand{\nump}[0]{d}

\begin{document}

\title{Dimension matters: precision and incompatibility in multi-parameter quantum estimation models}

\author{Alessandro Candeloro}\email{candeloa@tcd.ie}
\affiliation{School of Physics, Trinity College Dublin, Dublin 2, Ireland} 
\author{Zahra Pazhotan}
\affiliation{Quantum Technology Lab, Dipartimento di Fisica
{\em Aldo Pontremoli}, Universit\`a degli Studi di Milano, I-20133, 
Milano, Italy}
\author{Matteo G.A. Paris}\email{matteo.paris@fisica.unimi.it}
\affiliation{Quantum Technology Lab, Dipartimento di Fisica
{\em Aldo Pontremoli}, Universit\`a degli Studi di Milano, I-20133, 
Milano, Italy}
\affiliation{Istituto Nazionale di Fisica Nucleare, Sezione di Milano, 
I-20133, Milano, Italy.}
\begin{abstract}
We study the role of probe dimension in determining the bounds of precision and the level of incompatibility in multi-parameter quantum estimation problems. In particular, we focus on the paradigmatic case of unitary encoding generated by $\mathfrak{su}(2)$ and compare precision and incompatibility in the estimation of the same parameters across representations of different dimensions. For two- and three-parameter unitary models, we prove that if the dimension of the probe is smaller than the number of parameters, then simultaneous estimation is not possible (the quantum Fisher matrix is singular). If the dimension is equal to the number of parameters, estimation is possible but the model exhibits maximal (asymptotic) incompatibility. However, for larger dimensions, there is always a state for which the incompatibility vanishes, and the symmetric Cramér-Rao bound is achievable. We also critically examine the performance of the so-called asymptotic incompatibility (AI) in characterizing the difference between the Holevo-Cramér-Rao bound and the Symmetric Logarithmic Derivative (SLD) one, showing that the AI measure alone may fail to adequately quantify this gap. Assessing the determinant of the Quantum Fisher Information Matrix (QFIM) is crucial for a precise characterization of the model's nature. Nonetheless, the AI measure still plays a relevant role since it encapsulates the non-classicality of the model in one scalar quantity rather than in a matrix form (i.e., the Uhlmann curvature).
\end{abstract}
\maketitle
%

%
\section{Introduction}
Quantum metrology is a cornerstone of quantum technology, playing a crucial role in its advancement. Precise characterization of quantum systems is paramount for developing future quantum technologies, with numerous quantum sensors already finding applications in both academic research and industry. Examples include gravitational-wave detectors \cite{schnabel2010quantum}, quantum clocks \cite{RevModPhys.87.637}, and quantum imaging systems \cite{Brida2010,polino2020photonic}.
\par

Quantum parameter estimation forms the theoretical foundation of quantum metrology, whose goal is to achieve the highest possible precision in estimating a specific parameter of interest. One of its main results is the Quantum Cramér-Rao bound \cite{helstrom1969quantum,holevo2011probabilistic,braunstein1994statistical}, which serves as the lower bound on the precision achievable by all possible measurement strategies \cite{paris2009quantum}. In general, this bound is attainable with adaptive strategies \cite{Fujiwara_2006}, and determining the optimal measurement remains a standard challenge in quantum parameter estimation.
\par
However, distinctive quantum features emerge primarily in multi-parameter scenarios, where the role of incompatibility becomes fundamental \cite{albarelli2020perspective}. In principle, the single-parameter Quantum 
Cramér-Rao bound can be extended to a matrix inequality, known as the matrix Quantum Cramér-Rao bound. However, this bound is generally unattainable due to the lack of compatibility between optimal measurements for individual parameters \cite{ragy2016compatibility}. A scalar bound, first derived by Holevo, is instead attainable in multi-parameter scenarios and is known as the Holevo-Cramér-Rao bound \cite{holevo2011probabilistic}. Unlike the single-parameter case, where local measurements are sufficient to attain the optimal bound, the Holevo Cramér-Rao bound proves to be generally attainable only in the most general scenarios that involves collective measurements on an asymptotically large number of copies of the quantum state \cite{kahn2009local,yang2019attaining}.
\par
A gap exists between the Holevo-Cramér-Rao bound and the scalar version 
of the quantum Cramér-Rao bound, which can be interpreted as a measure of the asymptotic incompatibility within the quantum statistical model under study \cite{albarelli2019evaluating,conlon2022gap}. Recently, there has been interest in exploring the property of this gap, and an upper bound $\mathcal{R}_{\bmp}$ was proved which is relatively straightforward to compute, relying solely on the symmetric logarithmic derivatives and the quantum state \cite{carollo2019quantumness}: we will refer to $\mathcal{R}_{\bmp}$ as Asymptotic Incompatibility (AI) measure. The value of this upper bound has been investigated for simple systems, such as qubits \cite{razavian2020quantumness}, for more complex problems like quantum tomography of finite dimensional systems \cite{candeloro2021properties}, and in case of a first order phase transition \cite{di2022multiparameter}. However, its features are still not fully explored, either in the context of other estimation scenarios or in terms of its tightness with respect to the true gap. The aim of this paper is to jointly address  these two questions.
\par

A significant set of estimation problems involves unitary transformations, where parameters are encoded through the evolution of a specific parameter-dependent Hamiltonian operator \cite{bisio2010optimal, baldwin2014quantum, liu2015quantum,liu2017quantum}. To enhance precision and develop improved estimation protocols, strategies have been devised to encode parameters in systems exhibiting increasingly greater sensitivity. Examples include systems that exploit quantum correlations \cite{PhysRevA.78.032303, giovannetti2006quantum, giovannetti2011advances} or involve quantum phase transitions \cite{zan08,macieszczak2016dynamical,frerot2018quantum,garbe2020critical}, demonstrating enhancement over the optimal classical scaling. However, the role of a quantum system's size—where 'size' refers to the dimension of the probe's Hilbert space—has received less attention. Our goal is to explore this aspect, particularly whether the Hilbert space dimension of the probe affects the incompatibility of the quantum statistical model, using the asymptotic incompatibility as a witness.
\par
In this paper, we focus on a particular unitary encoding: the one generated 
by $\mathfrak{su}(2)$ algebra \cite{yang2022multiparameter}. This choice facilitates a fair comparison between probes of different dimension. Indeed, 
each dimension corresponds to a distinct representation of the 
$\mathfrak{su}(2)$ algebra, allowing us to assess and compare their incompatibility in the estimation of the same parameter across different representations.
This article is driven by two primary objectives. First, we explore how the AI measure is influenced by the dimensionality of the probe, specifically examining the conditions under which the AI measure equals zero, which denotes a fully compatible statistical model. Second, we assess the effectiveness of the AI measure in accurately quantifying the gap between the Holevo-Cramér-Rao bound and the SLD-based one. Our aim is to uncover potential limitations and weaknesses of the AI measure in certain parameter regimes, highlighting instances where it fails to correctly represent the gap's behavior, thereby overestimating the incompatibility of quantum statistical models. These objectives are complementary rather than contradictory; the AI measure acts as an upper bound on the gap, a fundamentally positive quantity. Thus, a small AI measure, as investigated in our first inquiry, should consistently and accurately mirror the gap's dynamics.
\par
The paper is structured as follows: in Sec. \ref{sec:incomp-qumetr} we provide a review of multiparameter metrology in the quantum scenario, by introducing the key quantities that we investigate in the following sections, namely the AI measure $\mathcal{R}_{\bmp}$ and the gap $\Delta_{\bmp}$; in Sec. \ref{sec:metro-unitary-encod} we introduce the estimation problem we are going to address and we derive some general formulae for the unitary estimation problem being considered; in Sec. \ref{sec:two-param-estim} we study the unitarily $\mathfrak{su}(2)$-encoded estimation and we derive general formula for the QFIM and Uhlmann for the the angle $\theta$ and the strength $B$ parameters; we then address specifically qubit and qudit estimation respectively in Sec. \ref{sec:two-param-estim-qubit} and Sec. \ref{sec:two-param-estim-qudit}; we finally study the three parameter estimation problem encoded via $\mathfrak{su}(2)$ algebra in Sec. \ref{sec:three-param-estim}; eventually, we summarize our results in Sec. \ref{sec:conclusions}.

\section{Incompatibility in multi-parameter metrology}
\label{sec:incomp-qumetr}

We consider a family of quantum state $\denp$ that depends on a set of $\nump$ parameters $\bm\lambda$. The goal of estimation theory is to devise the optimal procedure to estimate these parameters, and the most general way to describe a measurement is given in terms of positive operator-valued measurements (POVMs), a collection of positive semi-definite operator $\bm{\Pi}=\{\Pi_i\}_i$ such that $\sum_{i}\Pi_i = \mathbb{I}$. 
\par 
In this article, we study local estimation theory, in which the usual figure of merit to quantify the precision of a given measurement $\bm{\Pi}$ is the covariance matrix $\bm{V}(\hat{\bm\lambda})$ of an estimator $\hat{\bm\lambda}$. The fundamental result of classical estimation theory is that $\bm{V}(\hat{\bm\lambda})$ is lower bounded as 
\begin{equation}
	\bm{V}(\hat{\bm\lambda}) \geq \bm{\mathcal{F}}^{-1}_{\bmp}(\mathcal{P})\, ,
    \label{eq:CRB}
\end{equation}
where 
\begin{equation}
    \left[\bm{\mathcal{F}}_{\bmp}(\mathcal{P})\right]_{jk} = \sum_{i} \left[\partial_j\log p(i\vert\bm\lambda) \right]\,\left[\partial_k \log p(i\vert\bm\lambda)\right]\, p(i\vert\bm\lambda)
\end{equation}
is the classical Fisher Information matrix (FIM) of the probability distribution $\mathcal{P}=\{p(i\vert\bm\lambda)\}_i$. One can prove that the bound in Eq. \eqref{eq:CRB} tight at least in the asymptotic limit \cite{lehmann2006theory}. 
\par 
Before analysing the concept of quantum incompatibility in parameter estimation, it is important to note that a similar incompatibility notion is already present at the classical level. Specifically, it is possible for a set of parameters $\lambdaB$ to be not estimated simultaneously. Indeed, we have the following theorem, that connects the size of a random variable with the number of estimable parameters and the invertibility of the FIM.
\par
\begin{theorem}[Invertibility of FIM]
	Given a random variable $X$ with $n$ outcomes described by the probability distribution $\{p_i(\bm\lambda)\}_{i=1}^{n}$ that depends on $d$ parameters $\bm\lambda = \{\lambda_{1},...,\lambda_{d}\}$, than the corresponding FIM is not invertible if $n<d+1$.
\end{theorem}
The proof is reported in \ref{app:proof1}. The theorem simply indicates that if the number of outcomes is not large enough, i.e. less than the number of parameters plus 1, it is not possible to estimate all the parameters simultaneously, but we can only estimate functions of them. In other words, we must restrict ourselves to the estimation of $d'\leq n$ functions of the original parameters. This results in a notion of of \emph{classical estimation incompatibility}.
\par 
When we move to the quantum realm, probability distributions naturally emerge from quantum mechanics postulates. Indeed, the Born rule is
\begin{equation}
    p(i\vert \bm{\lambda}) = \Tr{\denM_{\bm{\lambda}} \Pi_i}.
\label{eq:bornrule}
\end{equation}
We can generalize the FIM to the quantum case, introducing the Quantum Fisher information matrix (QFIM) \cite{paris2009quantum}
\begin{equation}
    [\bm{\mathcal{Q}}_{\bmp}]_{jk} = \frac{1}{2}\Tr{\denp \{\SLDp{\lambda_j},\SLDp{\lambda_k}\}} 
\label{eq:qfimatrix}
\end{equation}
where the operator $L_{\lambda_i}$ is the symmetric logarithmic derivative (SLD) operator defined implicitly as the solution of the Lyapunov equation $2\partial_{\lambda_i} \denp = \{\SLDp{\lambda_i},\denp\}$. This matrix yields to the SLD-Cramer-Rao (SLD-CR) bound
\begin{equation}
	\bm{V}(\tilde{\bmp}^{(M)}) \geq \bm{\mathcal{Q}}_{\bmp}^{-1}
	\label{eq:SLD-crb}
\end{equation}
A scalar bound can also be introduced with respect to a certain weight matrix $\bm{W}$ of the parameters using the cost function \cite{albarelli2020perspective}
\begin{equation}
	\mathcal{C}[\bm{V}(\tilde{\bmp}^{(M)}),\bm{W}] = \tr{\bm{W}\bm{V}(\tilde{\bmp}^{(M)})}.
	\label{eq:cost-funct}
\end{equation}
and a scalar-SLD-CR bound can be straightforwardly derived
\begin{align}
	\mathcal{C}[\bm{V}(\tilde{\bmp}^{(M)}),\bm{W}] & \geq \tr{\bm{W}\bm{\mathcal{Q}}_{\bmp}^{-1}}  = \mathcal{C}^{\text{SLD}}[\denp,\bm{W}]
	\label{eq:scalar-SLD-crb}
\end{align}
To avoid any confusion, we stress here that we use $\tr{}$ to denote the trace on matrix, while $\Tr{}$ is used for the trace on density operators. 
\par
These bounds are in general not attainable, and this can be heuristically understood by the following minimal argument: in the case of a single parameter parameter $\lambda$, the bound is naturally a scalar bound and it is tight. The optimal measurement is given by the eigenprojectors of the SLD operator $\SLDp{\lambda}$. Considering a pair of parameters $\{\lambda_1,\lambda_2\}$, the two optimal POVMs, given by the corresponding SLD operators, may not commute, and thus may not be performed simultaneously.
The sufficient and necessary condition to have a tight bound in Eq. \eqref{eq:SLD-crb} can not be directly derived from the definition of the QFIM. As a matter of fact, one can define a tighter bound known as the Holevo-Cramer-Rao (HCR) bound \cite{holevo2011probabilistic}. This is obtained by the following minimization \cite{demkowicz2020multi} 
\begin{align}
    \mathcal{C}^{\text{H}}[\denp,\bm{W}] & =  \min_{\bm{V} \in \mathbb{S}^\nump, {\bm{X}} \in \mathcal{X}_{\bmp}} \left\{\tr{\bm{W} \bm{V}}{} \vert \bm{V} \geq \bm{Z}\left[{\bm{X}}\right]\right\}
\label{eq:holevo-crb}
\end{align} 
where $\mathbb{S}^\nump$ is the set of real symmetric $\nump$-dimensional matrices.
The $\nump\times\nump$ matrix $\bm{Z}\left[{\bm{X}}\right]$ is defined as
\begin{equation}
\bm{Z}\left[{\bm{X}}\right]_{ij} = \Tr{X_i X_j \denp}{} \quad i,j=1,..., \nump,
\end{equation}
while the collection of $\nump$ operators ${\bm{X}}$ belongs to the set 
\begin{align}
\mathcal{X}_{\bmp} = \{ & {\bm{X}} = (X_1,...,X_{\nump}) \vert X_i \in \mathcal{L}_h(\mathcal{H}) \hspace{5pt} \& \hspace{5pt} \Tr{X_i \partial_{\lambda_{j}}\denp} = \delta_{ij} \}
\end{align}
In quantum multi-parameter estimation, this bound is considered the most fundamental scalar bound since it can be proved to be equivalent to the most informative bound on the asymptotic model with collective measurements on a $M\to\infty$ number of copies of the state $\denp^{\otimes M}$ \cite{yang2019attaining}. General analytic solution for the HCR bound are very rare. Nonetheless, it can be numerically evaluated via semidefinite programming \cite{albarelli2019evaluating}. 
\par
Analytic solutions can be found when the HCR and SLD-CR bound can be proven equal. A simple condition for the saturation of the SLD-CR bound (both in the matrix and in the scalar case) is given in terms of the Uhlmann matrix
\begin{equation}
	\bm{\mathcal{D}}_{\bmp} = \frac{1}{2i} \Tr{\denp [{\bm{L}},{\bm{L}}^T]}.
\end{equation}
The condition $\mathcal{C}^{\text{H}}[\denp,\bm{W}] =\mathcal{C}^{\text{SLD}}[\denp,\bm{W}]$ is obtained if the {\em weak compatibility} condition is satisfied \cite{ragy2016compatibility}
\begin{equation}
	\bm{\mathcal{D}}_{\bmp} = 0.
	\label{eq:weakcomp}
\end{equation}
The terms {\em weak} refers to the fact d that, rather the commutation of all the SLDs, it is sufficient for the SLDs to commute on average only, i.e., on the support of $\denp$. In that case, the SLD-CR bound is tight, but we still stress that it might be attainable only in the asymptotic limit. Hence, the Uhlmann matrix has relevance only in the asymptotic limit, and in case Eq. \eqref{eq:weakcomp} is satisfied, the model is known as {\em asymptotically classical} \cite{albarelli2020perspective}.
\par 
In the opposite scenario, i.e. when Eq. \eqref{eq:weakcomp} is not satisfied, one may wonder how much is the gap between the SLD-CR bound and the tigh HCR bound.  One can prove that the normalized gap is upper bounded as \cite{carollo2019quantumness}
\begin{equation}
	\Delta_{\bmp} \equiv \Delta[\denp,\bm{W}] = \frac{\mathcal{C}^{\text{H}}[\denp, \bm{W}]- \mathcal{C}^{\text{SLD}}[\denp, \bm{W}]}{\mathcal{C}^{\text{SLD}}[\denp, \bm{W}]}   \leq \mathcal{R}_{\bmp}
\end{equation}
where we have defined the asymptotic incompatibility (AI) measure 
\begin{equation}
    \mathcal{R}_{\bmp} = \Vert \bm{\mathcal{Q}}_{\bmp}^{-1}\bm{\mathcal{D}}_{\bmp} \Vert_{\infty}
\end{equation}
and $\Vert A \Vert_{\infty}$ is the largest eigenvalue of $A$. A very important property of AI is that it is upper bounded as $\mathcal{R}_{\bmp} \leq 1$, and model with $\mathcal{R}_{\bmp} =1$ are known as maximally incompatible. From this upper bound, we conclude that the gap between the HCR and SLD-CR bound is at most the value of the SLD-CR bound. In other words, possible advantageous scaling in the HCR bound can be tested directly from the scaling of the SLD-CR bound. A further property of the AI is  its monotonicity with respect to quantum estimation sub-model $\mathcal{R}_{\tilde{\bmp}}$, i.e. the AI corresponding to the parametric model with the $\tilde{\nump} < \nump$ parameters $\tilde{\bmp} = \{\lambda_1,...,\lambda_{\tilde{\nump}}\}$ \cite{candeloro2021properties}.
\par
In the case of a pure model we have that the Holevo bound is exactly given as \cite{suzuki2016explicit}
\begin{equation}
    C^H[\denp,\bm{W}] = C^{\text{SLD}}[\denp,\bm{W}] + \Vert \bm{\mathcal{Q}}_{\bmp}^{-1}\bm{\sqrt{\bm{W}}\mathcal{D}_{\bmp}} \bm{\mathcal{Q}}_{\bmp}^{-1} \sqrt{\bm{W}}\Vert_1
\end{equation}
and thus we can identify the gap as
\begin{equation}
	\Delta_{\bmp} = \frac{\Vert \bm{\mathcal{Q}}_{\bmp}^{-1}\bm{\mathcal{D}}_{\bmp} \bm{\mathcal{Q}}_{\bmp}^{-1}\Vert_1}{C^{\text{SLD}}[\denp,\bm{W}] }
\end{equation}
We will use this equation in the following to evaluate how much the AI measure is overestimating the true gap in a pure state model.

\section{QFIM and Uhlmann matrix for unitary encoded parameters}
\label{sec:metro-unitary-encod}

Let us assume that the parameters are encoded via a unitary evolution $U_{\bm\lambda}$, i.e. $\vert \psi_{\bm\lambda} \rangle = U_{\bm\lambda} \vert \psi_{0}\rangle$. In this case, the QFIM $\bm{\mathcal{Q}}_{\bmp}$ and the Uhlmann matrix $\bm{\mathcal{D}}_{\bmp}$ are given as \cite{liu2015quantum}
\begin{align}
\mathcal{Q}_{ll'} & =  2 \langle \left\{\mathcal{H}_{l},\mathcal{H}_{l'}\right\} \rangle_{0} - 4 \langle \mathcal{H}_{l}\rangle_{0} \langle \mathcal{H}_{l'}\rangle_{0} \label{eq:qfimunitary} \\
\mathcal{D}_{ll'} & =  -2i \langle \left[\mathcal{H}_{l},\mathcal{H}_{l'}\right] \rangle_{0} \label{eq:imunitary}
\end{align}
where we have defined the hermitian operator $\mathcal{H}_l$ as 
\begin{equation}
\mathcal{H}_{l} = i (\partial_{l} U^{\dagger}) U 
\end{equation}
and where $\langle \bullet \rangle_0$ denotes the expectation value with respect to the initial state of the probe $\vert \psi_0\rangle$, while $\partial_l = \partial/\partial \lambda_l$.
\par 
The expression given in \eqref{eq:qfimunitary} and \eqref{eq:imunitary} relies on the operator $\mathcal{H}_{l}$. Despite a general closed formula is not easy to derive for $\mathcal{H}_l$, we can re-express it in a more suitable way. The starting point is the expression for the derivative of the exponential of an operator $A_{\alpha}$ that depends on a certain parameter $\alpha$ given as
\begin{equation*}
\partial_{\alpha}e^{A_{\alpha}} = \int_{0}^{1}ds e^{sA_{\alpha}}(\partial_{\alpha} A_{\alpha} ) e^{(1-s)A_{\alpha}}
\end{equation*}
In our case $U=e^{-itH_{\alpha}}$, we have that
\begin{align}
    \mathcal{H}_{\alpha} = i \sum_{n=0}^{+\infty} f_{n} H^{\times}_{\alpha}(\partial_{\alpha} H_{\alpha})^{n}
\label{eq:calH}
\end{align} 
with $f_{n} = (it)^{n+1}/(n+1)!$, and the superoperator $A^{\times}(\bullet) = [A,\bullet]$. If this last expression has some recursive structure, then we might be able to write $\mathcal{H}_{l}$ in a closed formula. We anticipate that this is our case.

\section{Two-parameters estimation of an $\mathfrak{su}(2)$ unitary encoding}
\label{sec:two-param-estim}
We consider the 2-parameter estimation problem
\begin{equation}
H_{B,\theta} = H = B (\cos\theta J_{x} + \sin\theta J_{z}) = B J_{\bm{n}_{\theta}}
\end{equation}
where $J_{x}$, $J_{y}$ and $J_{z}$ belongs to $\mathfrak{su}(2)$ Lie algebra and satisfy the following identities
\begin{align}
[J_{x},J_{y}] = i J_{z} \label{eq:spinx}\\
[J_{y},J_{z}] = i J_{x} \label{eq:spiny}\\
[J_{z},J_{x}] = i J_{y} \label{eq:spinz}
\end{align}
In the following, we are going to use the following notation $J_{\bm{n}_{\theta}} = \bm{n}_{\theta} \cdot \bm{J}$, where the vectors are defined as $\bm{n}_{\theta} = \left(\cos\theta,0,\sin\theta \right)$ and $\bm{J} = \left(J_{x},J_{y},J_{z}\right)$ is the vector of the generator operators.
\par
As a first step, we are going to evaluate the QFIM and Uhlmann matrix. We stress that we \emph{did not} specify the dimensionality of the Hilbert space yet, we have just considered that the generator of the encoding parameter satisfy the $\mathfrak{su}(2)$ Lie algebra. To evaluate $\mathcal{H}_{l}$ for $l=\theta,B$, we need to evaluate $H^{\times}_{l}(\partial_{l}H_{l})$. We start considering the paramater $\theta$. In this case
\begin{equation}
\partial_{\theta} H = B J_{\bm{n}'_{\theta}}
\end{equation}
with $\bm{n}'_{\theta}=(-\sin(\theta),0,\cos(\theta))$. Then we notice that 
\begin{align} 
    H^{\times}(J_{y}) = [H,J_{y}] = i B J_{\bm{n}_{\theta}'} = i \partial_{\theta} H
\end{align}
From that it follows 
\begin{align}
H^{\times} (\partial_{\theta} H) &= H^{\times} (-i H^{\times}(J_{y})) = -i H^{\times 2}(J_{y})
\end{align}
and recursively we obtain 
\begin{equation}
    H^{\times n} (\partial_{\theta} H) = H^{\times n} (-i H^{\times}(J_{y})) = -i H^{\times (n+1)}(J_{y})
\end{equation}
Eventually, the operator $\mathcal{H}_{\theta}$ can be written as
\begin{align}
\mathcal{H}_{\theta} & = i \sum_{n=0}^{+\infty} f_{n} H^{\times n} (\partial_{\theta}H) =  \sum_{n=0}^{+\infty} f_{n-1} H^{\times (n)} (J_{y}) -J_{y}
\end{align}
Since $f_{n-1} = i t^{n}/n!$, we have that the series yield to the exponential operator
\begin{align}
\mathcal{H}_{\theta} & = \left\{\exp{i t H^{\times}} - \mathbb{I} \right\} J_{y}  = \left\{\exp{i t B J^{\times}_{\bm{n}_{\theta}}} - \mathbb{I} \right\} J_{y}
\end{align}
where we have used that $H^{\times}(\bullet) = B J^{\times}_{\bm{n}_{\theta}}(\bullet)$. This expression can be further simplified using the properties of the $\mathfrak{su}(2)$ algebra, i.e. 
\begin{align*}
J^{\times}_{\bm{n}_{\theta}}(J_{y}) & = [J_{\bm{n}_{\theta}},J_{y}] = i J_{\bm{n}'_{\theta}} \\
J^{\times}_{\bm{n}_{\theta}}(J_{\bm{n}_{\theta}'}) & = [J_{\bm{n}_{\theta}},J_{\bm{n}_{\theta}'}] = - i J_{y}
\end{align*}
We see that the recursive application of the superoperator is closed, i.e. it remains in the subspace defined by the last two equation. Hence, we have that
\begin{equation}
J^{\times n}_{\bm{n}_{\theta}} (J_{y}) = \begin{cases}
J_{y} & n\,\, \text{even} \\
i J_{\bm{n}_{\theta}'} & n\,\, \text{odd} 
\end{cases}
\end{equation}
which means that 
\begin{align}
\mathcal{H}_{\theta} & = J_{y}(\cos Bt - 1) - J_{\bm{n}_{\theta}'} \sin Bt = 2  J_{\bm{n}_{1}} \sin \frac{Bt}{2}
\end{align}
where we have defined the vector $\bm{n}_{1}$ with unit norm
\begin{equation}
\bm{n}_{1} = \left(\cos \frac{Bt}{2} \sin\theta,-\sin\frac{Bt}{2}, -\cos\frac{Bt}{2}\cos\theta\right)
\end{equation}
Regarding the parameter $B$, the calculations are much easier and the result is easily obtained as
\begin{equation}
\mathcal{H}_{B} = -t J_{\bm{n}_{\theta}}.
\end{equation}
From the expression derived in the previous subsection, the QFIM and Uhlmann matrix for a pure state model are straightforward. Indeed, we have that the QFIM elements are 
\begin{align}
\mathcal{Q}_{BB} & =  4 t^{2} (\langle J_{\bm{n}_{\theta}}^{2}\rangle_{0} - \langle J_{\bm{n}_{\theta}}\rangle^{2}_{0})\\
\mathcal{Q}_{\theta \theta} & =  16  \sin^{2}\frac{Bt}{2}  (\langle J_{\bm{n}_{1}}^{2}\rangle_{0} - \langle J_{\bm{n}_{1}}\rangle^{2}_{0}) \\
\mathcal{Q}_{B\theta} & = -4t  \sin \frac{Bt}{2}  \left( \langle \{J_{\bm{n}_{1}},J_{\bm{n}_{\theta}}\}\rangle_{0} - 2\langle J_{\bm{n}_{1}}\rangle_{0}\langle J_{\bm{n}_{\theta}}\rangle_{0}\right)
\end{align}
Regarding the Uhlmann matrix, the anti-commutator can be explicitly evaluated as
\begin{align}
[\mathcal{H}_{B},\mathcal{H}_{\theta}] & = -2\, i\, t  \sin\!\frac{Bt}{2}  J_{\bm{n}_{2}}    
\end{align}
where we used the fact that
\begin{equation}
[J_{\bm{n}_{\theta}},J_{\bm{n}_{1}}] = i J_{\bm{n}_{2}}
\end{equation}
with $\bm{n}_{2}=\bm{n}_{\theta}\times \bm{n}_{1}$. From this,  the Uhlmann matrix naturally follows as 
\begin{equation}
\mathcal{D}_{\theta B} = 4 t   \sin\frac{Bt}{2}  \langle J_{\bm{n}_{2}}\rangle_{0}
\end{equation}
with $\bm{n}_{2}$ that can be explicitly written as
\begin{align}
\bm{n}_{2} & = \bm{n}_{\theta} \times \bm{n}_{1} = \left(\sin\frac{Bt}{2}\sin\theta,\cos\frac{Bt}{2},-\sin\frac{Bt}{2}\cos\theta\right)
\end{align}
We have now all the ingredients to evaluate the AI measure that for two parameters has the following form
\begin{equation}
\mathcal{R}_{\bmp} = \sqrt{\frac{\det \bm{\mathcal{D}}_{\bmp}}{\det \bm{\mathcal{Q}}_{\bmp}}}
\end{equation}
The determinant of the Uhlmann matrix is
\begin{equation}
\det \bm{\mathcal{D}}_{\bmp} = 16 t^{2} \sin^{2}\frac{Bt}{2} \langle J_{\bm{n}_{2}}\rangle_{0}^{2}
\end{equation}
while the determinant of the QFIM is
\begin{align}
\det & \bm{\mathcal{Q}}_{\bmp} = 16 t^{2} \sin^{2}\frac{Bt}{2} \bigg ( 4\Delta_0[J_{\bm{n}_\theta}]\Delta_0[J_{\bm{n}_1}] -
\left( \langle \{J_{\bm{n}_{1}},J_{\bm{n}_{\theta}}\}\rangle_{0} - 2\langle J_{\bm{n}_{1}}\rangle_{0}\langle J_{\bm{n}_{\theta}}\rangle_{0}\right)^{2} \bigg)
\label{eq:detQ}
\end{align}
where we have define the variance that $\Delta_0[J_{\bm{n}}] = \langle J^2_{\bm{n}} \rangle_0-\langle J_{\bm{n}}\rangle_0^2$.
The AI then becomes
\begin{equation}
\mathcal{R}_{\bmp} =  \left(\frac{\langle J_{\bm{n}_{2}}\rangle_{0}^{2}}{4\Delta_0[J_{\bm{n}_\theta}]\Delta_0[J_{\bm{n}_1}] -  \bigg[ \langle \{J_{\bm{n}_{1}},J_{\bm{n}_{\theta}}\}\rangle_{0} - 2\langle J_{\bm{n}_{1}}\rangle_{0}\langle J_{\bm{n}_{\theta}}\rangle_{0}\bigg]^{2}}\right)^{\frac12}
\end{equation}
We stress that this formula is independent on the size of the Hilbert space. This makes it an ideal candidate for investigating how the initial state and the dimensionality of the Hilbert space influecence the incompatibility of the parametric model.
\par
Before delving into study of this quantity in the general case, it is already interesting to evaluate it for the case of $N=2$, which already exhibits interesting features.

\subsection{Two parameter estimation for a qubit}
\label{sec:two-param-estim-qubit}

The qubit case corresponds to the fundamental representation of the $\mathfrak{su}(2)$ Lie algebra, for which we have a closed expressions for both the anticommutator
\begin{equation}
    \{J_{\bm{n}_{\alpha}},J_{\bm{n}_{\beta}}\} = \frac{1}{2} \bm{n}_{\alpha}\cdot \bm{n}_{\beta},
\end{equation}
This yields to 
\begin{align}
	\{J_{\bm{n}_{1}},J_{\bm{n}_{\theta}}\} =\frac{1}{2} \bm{n}_{\theta}\cdot\bm{n}_{1}\mathbb{I}_{2}  = 0
\end{align}
since from direct calculation $\bm{n}_{\theta}\cdot \bm{n}_{1}=0$. Moreover, we have that
\begin{equation}
J^{2}_{\bm{n}} = \frac{1}{4} \vert \bm{n} \vert ^{2} \mathbb{I}_{2} = \frac{1}{4} \mathbb{I}_{2}
\end{equation}
This means that the expectation values are $\langle J_{\bm{n}}^{2}\rangle_{0} = 1/4$ and $
\langle J_{\bm{n}}\rangle_{0} = \bm{n} \cdot \bm{r}_{0}/2$ with arbitrary norm one vector $\bm{n}$. It follows that the QFIM elements are given as
\begin{align}
\mathcal{Q}_{BB} & =   t^{2} \left[1 - (\bm{n}_{\theta}\cdot \bm{r}_{0})^{2}\right] \\
\mathcal{Q}_{\theta \theta} & =  4  \sin^{2}\frac{Bt}{2}\,  \left[1 - (\bm{n}_{1}\cdot \bm{r}_{0})^{2}\right]\\
\mathcal{Q}_{B\theta} & = 2t  \sin\frac{Bt}{2}   (\bm{n}_{1}\cdot \bm{r}_{0})(\bm{n}_{\theta}\cdot \bm{r}_{0}) 
\end{align}
while the IM is 
\begin{equation}
\mathcal{D}_{\theta B} = 2 t   \sin\frac{Bt}{2}\, \bm{n}_{2}\cdot \bm{r}_{0}.
\end{equation}
The AI also simplifies as
\begin{align}
\mathcal{R}_{\bmp} &  = \sqrt{\frac{(\bm{n}_{2}\cdot \bm{r}_{0})^{2}}{1-(\bm{n}_{1}\cdot \bm{r}_{0})^{2}-(\bm{n}_{\theta}\cdot \bm{r}_{0})^{2}}} 
\end{align}
To study the compatibility condition, we define $\bm{n}_{2}\cdot \bm{r}_{0} = \cos\varepsilon$. Then, the Uhlmann matrix is equal to zero when $\varepsilon = \pi/2$. However, in this limit, the QFIM is singular. Indeed, the decomposition of $\bm{r}_{0}$ in two component, one parallel and one parallel to $\bm{n}_{2}^{\perp}$, leads us to
\begin{equation}
\bm{r}_{0} = \cos\varepsilon\, \bm{n}_{2} + \sin\varepsilon\, \bm{n}_{2}^{\perp}
\end{equation}
Given that the vector lives in a 3D dimensional space, the perpendicular component can be expanded as
\begin{equation}
\bm{n}_{2}^{\perp} = \cos\phi\, \bm{n}_{\theta} + \sin\phi\, \bm{n}_{1}
\end{equation}
with arbitrary $\phi$. Then, since $\bm{n}_{1}\cdot \bm{n}_{2}=\bm{n}_{\theta}\cdot \bm{n}_{2}=0$, we have that
\begin{align}
\bm{n}_{1}\cdot \bm{r}_{0} = \sin\varepsilon\, \bm{n}_{1}\cdot \bm{n}_{2}^{\perp} \\
\bm{n}_{\theta}\cdot \bm{r}_{0} = \sin\varepsilon\, \bm{n}_{\theta}\cdot \bm{n}_{2}^{\perp}
\end{align}
from which we can evaluate 
\begin{equation}
(\bm{n}_{1}\cdot \bm{r}_{0})^{2}+(\bm{n}_{\theta}\cdot \bm{r}_{0})^{2} =\sin^{2}\varepsilon
\end{equation}
It follows that 
\begin{equation}
\mathcal{R}_{\bmp} =  \sqrt{\frac{\cos^{2}\varepsilon}{1-\sin^{2}\varepsilon}} = 1
\end{equation}
We conclude that the AI measure always equals the maximum value, independently of the parameters values and the initial pure state. It follows that, when the Uhlmann matrix is zero, the determinant of the QFIM is also zero, and the limit yields to a maximal AI measure. Physically, a non-invertible QFIM implies that the two parameters are not independent, and this is why their optimal estimation can not be achieved simultaneously. In other words,  only a (non)-linear combination of the parameter can be estimated.\footnote{This can be seen considering a quantum statistical model $\varrho_{\bmp}$ whose QFIM is $\bm{\mathcal{Q}}_{\bmp}$. Then, we have that $\bm{\mathcal{Q}}_{\bmp} \geq \bm{\mathcal{F}}_{\bmp} \geq 0$, where $\bm{\mathcal{F}}_{\bmp}$ is the FIM of an arbitrary measurement. Then, if $\det \bm{\mathcal{Q}}_{\bmp} =0$ this implies that $\det \bm{\mathcal{F}}_{\bmp} = 0$ for any arbitrary measurements. This means that the parameters of the quantum statistical model can not be simultaneously estimated because they are not independent, regardless of the measurement procedure we perform.} We recall that we named this form of non-simultaneous estimation  as "classical incompatibility". This is to underline the following fact: that a strictly positive $\mathcal{R}_{\bmp}$ does not imply that $\bm{\mathcal{D}}_{\bmp} \neq 0$. Insteda, two possibilities arise: 1) $\det\bm{\mathcal{D}}_{\bmp} \neq 0$, and the quantum statistical model is not asymptotically classical; 2) $\det\bm{\mathcal{D}}_{\bmp} = 0$ and $\det\bm{\mathcal{Q}}_{\bmp} = 0$ yet their ratio's limit is finite. In this case, the model is asymptotically classical, but the simultaneous estimation of the parameter is not possible given their functional dependence. Therefore, relying only on $\mathcal{R}_{\bmp}$ may not conclusively determine the asymptotic compatibility of the quantum statistical model. 
\par
\begin{figure*}
  \includegraphics[width=\textwidth]{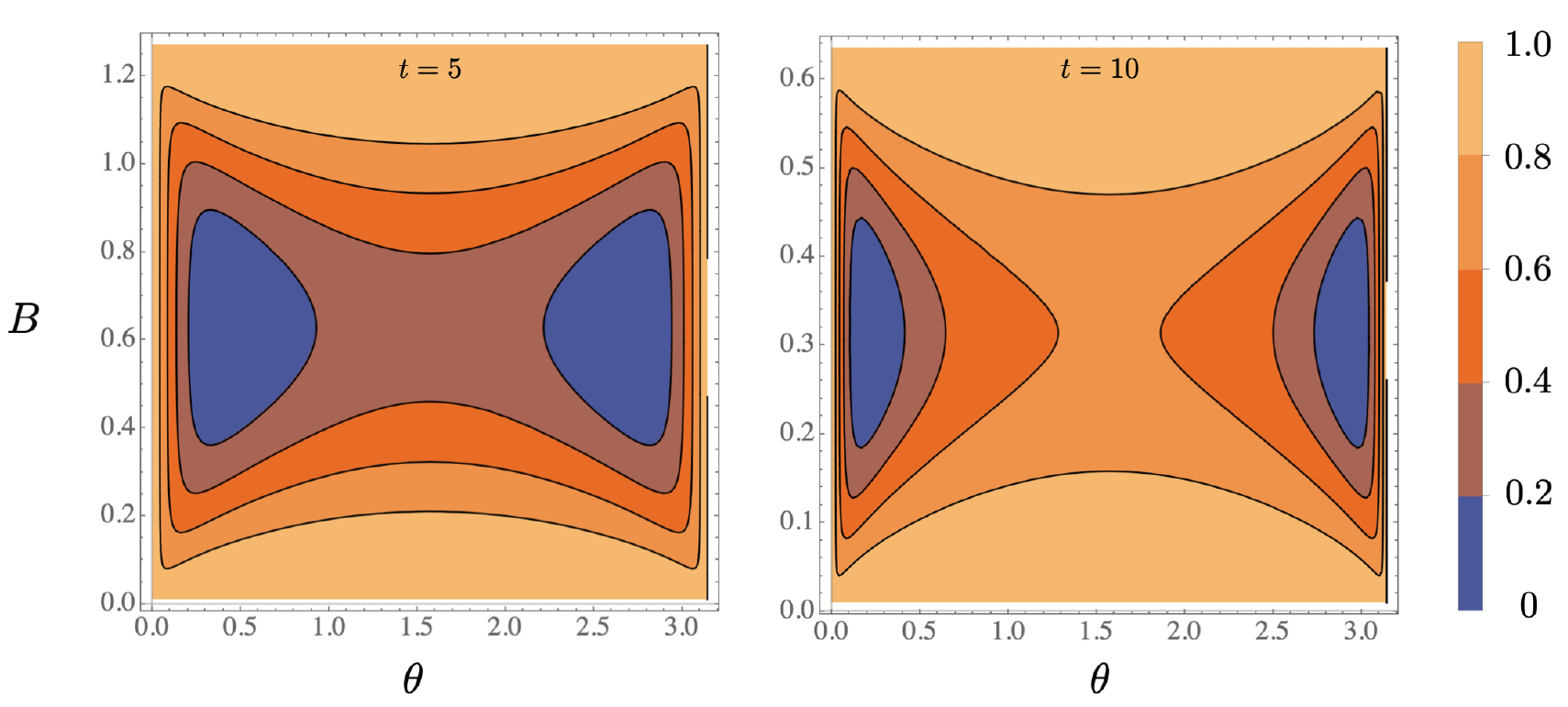}
  \caption{Contour plot of $T(\theta,B)=\mathcal{R}_{\bmp}-\Delta_{\bmp}$ for two parameter and for a qubit. Left panel: $t=5$; right panel: $t=10$. We have fixed the initial state as the superposition of the two eigenstate of $\sigma_{z}$, i.e. $1/\sqrt{2}(\vert 1\rangle + \vert 0)$. Notice that the two plot have different range for $B$. The reasons is that the figure of merit $T(\theta,B)$ is periodic in $B$, hence we have reported only the first period for the two times considered.} 
  \label{fig:comparison1}
\end{figure*}

 
Furthermore, as we observed in the introduction, the Holevo bound for pure state is given by $C^{H} = C^{S}(1+ \Delta_{\bmp})$ and we know that it is bounded by 
$C^{H} \leq  C^{S} (1+ \mathcal{R}_{\bmp})$. We thus may ask how much the AI measure is a good measure of the gap. Indeed, this can me by comparing the following quantities
\begin{equation}
\frac{C^{H} - C^{S}}{C^{S}} = \Delta_{\bmp} \leq \mathcal{R}_{\bmp} 
\end{equation}
In Fig. \ref{fig:comparison1} we report the positive function $T(\theta,B) = \mathcal{R}_{\bmp}-\Delta_{\bmp}$, representing the gap between the upper-bound on $\Delta_{\bmp}$ and its true value. When the function approaches $0$, the AI measure represents a good reliable bound for the gap. Conversely, when $T(\theta,B)$ approaches $1$, then the AI measure proves to be maximally overestimating the gap between $C^S$ and $C^H$. Here we see that there is a range of parameter for which the AI measure $\mathcal{R}_{\bmp}$ fails to capture the behaviour of $\Delta_{\bmp}$. We notice that this specifically happens for values of the parameters close to the values that makes the QFIM not invertible. Furthermore, we notice that as $t$ increase, the region of the contour plot where $T(\theta,B)$ is small constantly shrinks, another indication that the AI $\mathcal{R}_{\bmp}$ does not always succeed in describing the gap between the HCR bound and the SLD-CR bound.
\subsection{Two parameter estimation for a qudit}
\label{sec:two-param-estim-qudit}

We now consider larger finite-dimensional representation of $\mathfrak{su}(2)$. In particular, we consider irreducible representation of spin $s$ that corresponds to dimension $N=2s+1$. In this case, the analytic-closed evaluation the QFIM and Uhlmann matrix for a \emph{general} pure probe is not easy to derive. For this reason we consider a specific initial state of the form
\begin{equation}
\vert \psi_{0}\rangle = \frac{\cos\alpha\,\vert J\rangle_{J}+e^{i\phi}\sin\alpha\,\vert - J \rangle_{J}}{\sqrt{2}}
\label{eq:initialprobe}
\end{equation}
where $\vert \pm J\rangle_{J}$ are the eigenvectors with the largest and smallest eigenvalues of $J_{z}$. For single parameter estimation, this state corresponds to the optimal probe in the unitary model $e^{-i\theta H}$ whose generator's eigenstates $\vert J\rangle$ and $\vert - J\rangle$ corresponds to the maximum and minimum eigenvalues respectively \cite{seveso2020quantum}. We conjecture that this state might have a similar role in the multi-parameter setting, even though the generators for each parameter do not commute and thus a unique pair of eigenstates can not be determined. For a discussion on optimal probe in the quantum multi-parameter metrology, we refer the reader to \cite{albarelli2022probe}.
\par 
For $N>3$ (the expression for $N=3$ are slightly different, but the results on the estimation are the same) we have that the QFIM elements are
\begin{align*}
\mathcal{Q}_{BB} & = (N-1)\,t^{2} \left[\cos^{2}\theta+(N-1)\sin^{2} 2\alpha\, \sin^{2} \theta \right] \\
\mathcal{Q}_{\theta \theta} & = 4 (N-1) \sin^{4} \frac{Bt}{2}\bigg\{ 1 +
\cot^{2}\frac{Bt}{2} \left[(N-1)\cos^{2} \theta \sin^{2}2\alpha+\sin^{2}\theta \right]  \bigg\} \\
\mathcal{Q}_{\theta B} & = (N-1)\,\frac{t^{2}}{4} \left[N-3-(N-1) \cos 4\alpha \right]\,\sin Bt\,  \sin2\theta
\end{align*}
while the Uhlmann matrix element is 
\begin{equation}
\mathcal{D}_{\theta B} = -2t\, (N-1)\, \cos 2\alpha\, \cos\theta\, \sin^{2} \frac{Bt}{2}.
\end{equation}
We easily see that the only conditions such that the AI $\mathcal{R}_{\bmp}$ is zero is that $\alpha =\pi/4$, independently of $\phi$. Differently from the qubit case, it is straightforward to check that in this case the determinant of the QFIM is not zero and thus $\mathcal{R}_{\bmp}=0$. This means that we have found a initial probe $\vert \psi_0\rangle$ that yields to an asymptotic classical model where the two parameters can be simultaneously estimated.
\par 
A natural question arises at this point: is there a cost associated with reducing the quantum noise, i.e., selecting the probe with the minimal $\mathcal{R}_{\bmp}$? To address this question, we need to compare the optimal bound for different estimation strategies in different dimensions $N$. As discussed earlier, for $\alpha=\pi/4$, the optimal bound is given by the QFIM. In order to compare two different estimation in two different dimensions, say in dimension $N$ and $M$ we introduce the figure of merit
\begin{equation}
    \Gamma\left[\bm{\mathcal{Q}}_{\bmp}^{(N)},\bm{\mathcal{Q}}_{\bmp}^{(M)}\right] = \text{Tr}\left[\bm{\mathcal{Q}}_{\bmp}^{(N)} {\bm{\mathcal{Q}}_{\bmp}^{(M)}}^{-1}\right],
\label{eq:fig-of-merit-scaling}
\end{equation}
in a similar fashion to the quantity introduced in \cite{gill2000state}. In our case, for the optimal state $\alpha=\pi/4$ we have that
\begin{align}
    \Gamma[\bm{\mathcal{Q}}_{\bmp}^{(N)},\bm{\mathcal{Q}}_{\bmp}^{(2)}] \propto (N-1)^2
\end{align}
from which we conclude that the for the optimal state $\alpha=\pi/4$, the scaling in the QFIM is $N^{2}$. We can also show that the same scaling is obtained for any values of $\alpha$, meaning that the particular choice that minimise the quantum noise is not detrimental for the optimal scaling in the QFIM. Hence, not only we found a probe that simultaneously reduce the quantum noise as quantified by $\mathcal{R}_{\bmp}$, but we also obtain an $N^2$ scaling in the QFIM. We conclude that probe with larger dimension not only help in reducing the quantum noise but also improve the attainable bound in the asymptotic limit with a scaling proportional to the square of the dimension.
\par
\begin{figure*}
  \includegraphics[width=\textwidth]{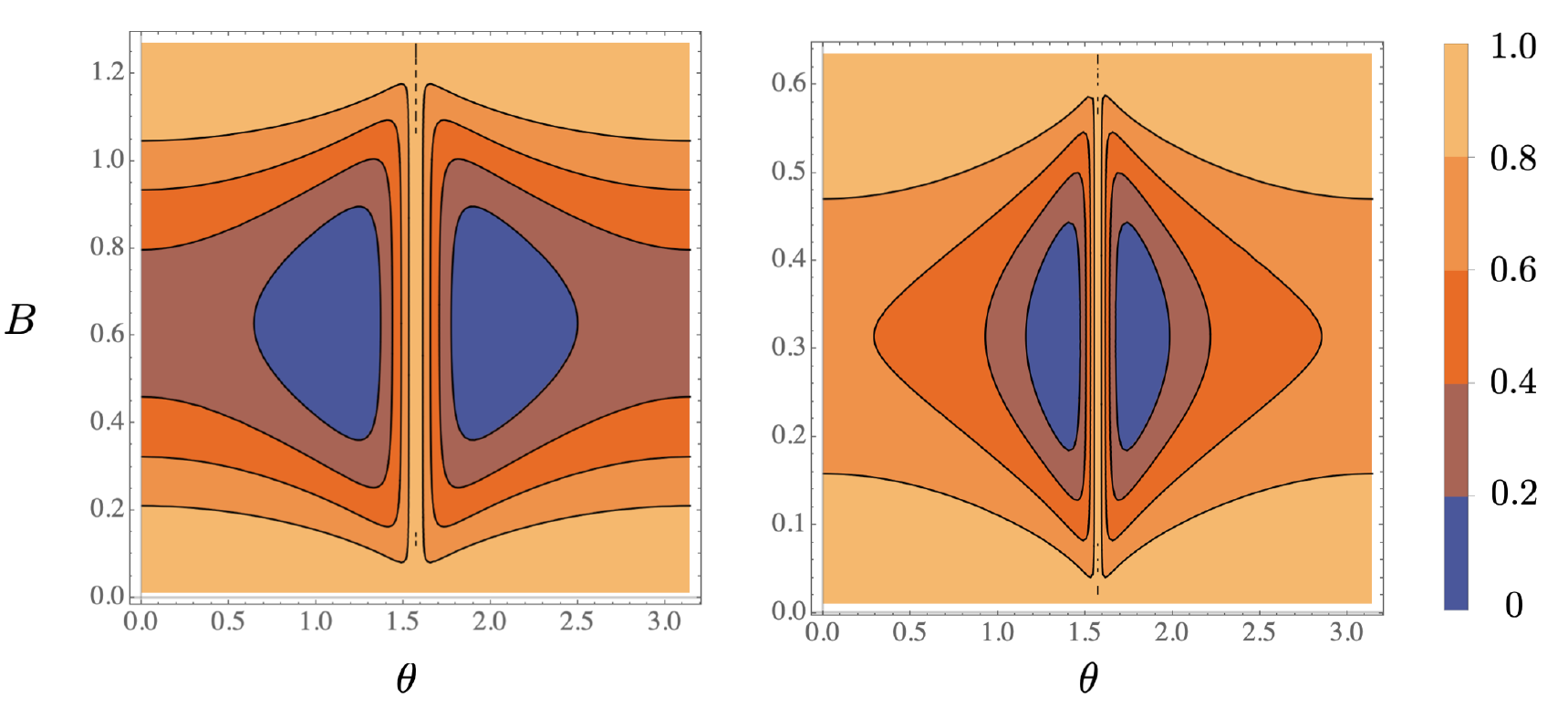}
  \caption{Contour plot  of $T(\theta,B) = \mathcal{R}_{\bmp}-\Delta_{\bmp}$ for $N=4$, and $\alpha=\pi/2$. Left panel: $t=5$; right panel $t=10$. Notice that the two plot have different range for $B$. The reasons is that the figure of merit $T(\theta,B)$is periodic in $B$, hence we have reported only the first period for the two times considered. The more the function approaches $1$, the more the AI measure is a good measure for the difference between the Holevo and the scalar bound.}
  \label{fig:comparison2}
\end{figure*}

The last step in our analysis is again to evaluate whether the AI measure $\mathcal{R}_{\bmp}$ is a faithful figure in measuring the gap between the HCR bound and SLD-CR bound. Similarly to what we did for the qubit, we can plot the same figure of merit $T(\theta,B)$ for $\alpha\neq\pi/4$, since in this case everything collapse to zero. The results are reported in  Fig. \ref{fig:comparison2}, and comparing the figure with the qubit case, we can observe a similar behaviour: there are regions where the AI measure is not a good estimate of the gap; and as time increase, the region in which the AI is a good measure shrinks, denoting that the AI is not always a good measure for the gap between the HCR bound and SLD-CR bound.

\section{Three-parameters estimation of an $\mathfrak{su}(2)$ unitary encoding}
\label{sec:three-param-estim}
After having studied the two-parameter problem in the previous section, it is now time to address the compatibility of the three-parameter problem for an $\mathfrak{su}(2)$ unitary encoding. In this case, the generator is given as 
\begin{equation}
H_{B,\theta,\varphi} = H = B J_{\bm{n}^{(3)}_{\theta}}
\end{equation}
where $J_{x}$, $J_{y}$ and $J_{z}$ belongs to $\mathfrak{su}(2)$ satisfy \eqref{eq:spinx}-\eqref{eq:spinz}. In the following, we are going to use the following notation $J_{\bm{n}^{(3)}_{\theta}} = \bm{n}_{\theta} \cdot \bm{J}$, where the vector is defined as $$\bm{n}^{(3)}_{\theta} = \left(\cos\theta \cos\varphi,\cos\theta\sin \varphi,\sin\theta \right)$$ and $\bm{J} = \left(J_{x},J_{y},J_{z}\right)$. 
\par 
With the same line of reasoning of the previous section, we are going to evaluate the QFIM and IM. Againg, we stress that we \emph{did not} specify the dimensionality of the Hilbert space yet. We start considering the paramater $\theta$. In this case
\begin{equation}
\partial_{\theta} H = B J_{\bm{n}^{(3)}_{\theta'}}
\end{equation}
with $\bm{n}^{(3)}_{\theta'}=(-\sin\theta\cos\varphi,-\sin\theta\sin\varphi,\cos\theta)$. Then, defining $\bm{n}_{0}=(-\sin\varphi,\cos\varphi,0)$ and after some algebra one also finds that 
\begin{equation}
    H^{\times}(J_{\bm{n}_{0}}) = [H,J_{\bm{n}_{0}}]  = i B J_{\bm{n}^{(3)}_{\theta'}} = i \partial_{\theta} H
\end{equation}
Then, with the exact same passages as in the case of two parameters we have that 
\begin{align}
\mathcal{H}_{\theta} = 2  \sin \frac{Bt}{2}\, J_{\bm{n}_{1}^{(3)}}
\end{align}
where we have defined the vector $\bm{n}^{(3)}_{1}$ with unit norm
\begin{align}
\bm{n}^{(3)}_{1} =&  \bigg( \sin\frac{Bt}{2} \sin\varphi + \cos\frac{Bt}{2}\sin\theta\cos\varphi, \notag \\ &
 -\sin\frac{Bt}{2}\cos\varphi+\cos\frac{Bt}{2}\sin\theta\sin\varphi,
-\cos\frac{Bt}{2}\cos\theta\bigg)
\end{align}
We now move to the parameter $\varphi$. In this case
\begin{equation}
\partial_{\varphi}H = B \cos\theta J_{\bm{n}_{\varphi'}}
\end{equation}
with $\bm{n}_{\varphi'} = (-\sin\varphi,\cos\varphi,0)$. We notice that
\begin{align}
H^{\times}(J_{\bm{n}_{\theta'}^{(3)}}) & = B[J_{\bm{n}_{\theta}^{(3)}},J_{\bm{n}_{\theta'}^{(3)}}] = -i B J_{\bm{n}_{\varphi'}} = -i \sec\theta\partial_{\varphi}H
\label{eq:Htimesvarphi}
\end{align}
The algebra in the derivation of $\mathcal{H}_\varphi$ resembles the one for the parameter $\theta$, but there are a few different passages, that we report in \ref{app:threeparam}. The final result is simply given as
\begin{align}
\mathcal{H}_{\varphi} & =  2  \sin\frac{Bt}{2} J_{\bm{n}^{(3)}_{2}}
\label{eq:Hvarphi}
\end{align}
where we have defined the vector $\bm{n}^{(3)}_{2}$ with unit norm as
\begin{align}
\bm{n}^{(3)}_{2} = & \bigg( \cos\frac{Bt}{2}\sin\varphi - \sin\frac{Bt}{2}\sin\theta\cos\varphi, \notag \\
& -\cos\frac{Bt}{2}\cos\varphi-\sin\frac{Bt}{2}\sin\theta\sin\varphi, 
\sin\frac{Bt}{2}\cos\theta\bigg)
\label{eq:n32vector}
\end{align}
Regarding the parameter $B$, the calculations are much easier, and we obtain that 
\begin{equation}
\mathcal{H}_{B} = -t J_{\bm{n}^{(3)}_{\theta}}
\end{equation}
We can now evaluate the QFIM and Uhlmann matrix, derived in Eq. \eqref{eq:qfimunitary} and Eq. \eqref{eq:imunitary}. The resulting expressions of QFIM elements are not particularly insightful, and thus we do not report them here. Instead, the Uhlmann matrix elements are given after some algebra and considering the fact that $\bm{n}^{(3)}_{1} = \bm{n}_{2}^{(3)} \times \bm{n}^{(3)}_{\theta}$. Eventually the IM elements are given by 
\begin{align}
\mathcal{D}_{B,\theta} &= -4 t \sin\frac{Bt}{2} \langle J_{\bm{n}_{2}^{(3)}} \rangle_{0} \\
\mathcal{D}_{B,\varphi} &= 4 t \sin\frac{Bt}{2} \langle J_{\bm{n}_{1}^{(3)}} \rangle_{0} \\
\mathcal{D}_{\theta,\varphi} &= 8  \sin^{2}\frac{Bt}{2} \langle J_{\bm{n}_{\theta}^{(3)}} \rangle_{0} 
\end{align}
It is not clear from these equations in which scenario we can satisfy the weak compatibility condition. However, thanks to the property of the AI measure, it is sufficient to have $\mathcal{R}_{\bmp}  = \Vert i \bm{\mathcal{Q}}_{\bmp}^{-1}\bm{\mathcal{D}}_{\bmp} \Vert_{\infty} = 0$ and determinant of $\bm{\mathcal{Q}}_{\bmp}$ different from zero in order to have an asymptotic classical model. We are going to study this equation for different dimension $N$ of the Hilbert space and for different representation of the $\mathfrak{su}(2)$.
\begin{figure*}
\centering
  \includegraphics[width=\textwidth]{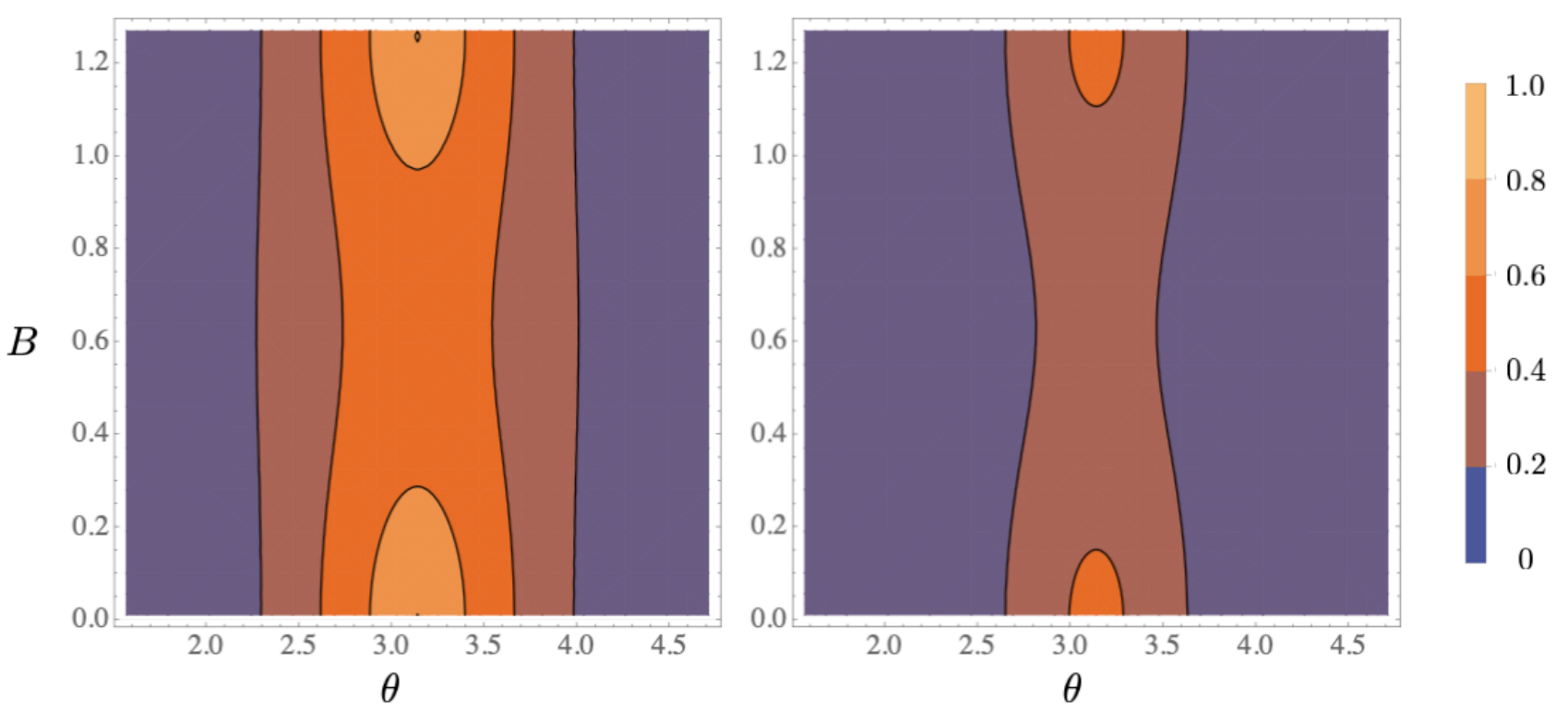}
  \caption{Contour plot of $T(\theta,B)=\mathcal{R}_{\bmp}-\Delta_{\bmp}$ for $t=5$, $N=4$ for the probe given by \eqref{eq:initialprobe}. Left plot: $\alpha=3\pi/5$; Right plot: $\alpha = 2\pi/3$.} 
  \label{fig:comparison3}
\end{figure*}
\par
We observe that in the case $N=2$, which corresponds to a qubit, it is already known that the Fisher Information Matrix (FIM) is not invertible. From a straightforward calculation, one can also deduce that the Quantum Fisher Information Matrix (QFIM) is not invertible. Consequently, the model is degenerate for all parameter values, implying that there are only two independent parameters instead of three.
\par 
Looking at the scenario with $N\geq 3$, a general expression for an arbitrary initial probe has no easy and compact expression. For that reason and with the same aim of the previous section, we consider the same initial probe given in Eq. \eqref{eq:initialprobe} to study the value of the AI measure, its faithfulness with respect to the true gap between the HCRb and the SLD-CRb, and finally a possible scaling in the QFIM as a function of $N$.
\par 
For $N=3$ it is not surprising to find a similar behaviour compared to the qubit case and two parameters. In this case, the AI measure is $\mathcal{R}_{\bmp}=1$ for all the values of $\alpha$ and $\phi$. However, as in the previous case, the weak compatibility condition in Eq. \eqref{eq:weakcomp} is satisfied for the same values that make the QFIM not invertible. This is another indication that we need to use of the AI measure with caution when having singular multiparameter estimation problems. Indeed, the asymptotic compatibility must be assesse by studying both the AI measure and the determinant of $\bm{\mathcal{Q}}_{\bmp}$. This is still easier than looking at each matrix element of elements $\bm{\mathcal{D}}_{\bmp}$.
\par
For $N\geq 4$, a straightforward calculation shows that the model asymptotic incompatibility measure is simply given as
\begin{equation}
\mathcal{R}_{\bmp} = \vert\cos 2\alpha\vert
\end{equation} 
from which we conclude that
\begin{equation}
    \mathcal{R}_{\bmp} = 0 \iff \alpha=\frac{\pi}{4} + \frac{\pi}{2}k
\end{equation}
with $k\in\mathbb{Z}$. Furthermore, for these values the QFIM is invertible and hence there is no singularity issue.
\par
We then investigate the scaling of the QFIM for the optimal probe given by $\alpha=\pi/4$, in order to check if there is a trade-off between incompatibility and the optimal scaling. We use the figure of merit defined in Eq. \eqref{eq:fig-of-merit-scaling}, and similarly to the previous case involving two parameters, we can show that 
\begin{equation}
    \Gamma[\bm{\mathcal{Q}}_{\bmp}^{(N)},\bm{\mathcal{Q}}_{\bmp}^{(2)}] \propto N^2
\end{equation}
for all the values of $\alpha$, including also the optimal case. We conclude that, by considering the probe in Eq.\eqref{eq:initialprobe} embedded in higher dimensional Hilbert space, we are able to reduce the quantum noise coming from the measurement incompatibility, quantified by the gap $\Delta_{\bmp}$ and the AI measure $\mathcal{R}_{\bmp}$, without loosing the optimal scaling. Even more, probes in higher dimensional Hilbert space prove to have a better metrological power, reflected in the quadratic scaling of $\Gamma[\bm{\mathcal{Q}}_{\bmp}^{(N)},\bm{\mathcal{Q}}_{\bmp}^{(2)}]$.
\par
Finally, we evaluate the gap between the AI measure and the true gap to assess the faithfulness of $\mathcal{R}_{\bmp}$. In Fig. \ref{fig:comparison3} , we plot $T(\theta,B)$ for the state given in \eqref{eq:initialprobe} for different value of $\alpha$. Again, there are large regions in which the AI measure overestimate the true gap. As expected, we also notice that the for values of $\alpha$ corresponding to smaller $\mathcal{R}_{\bmp}$, the region in which this happens is reduced. This corresponds to the fact that when $\mathcal{R}_{\bmp}$ is small, than deviation with respect to $\Delta_{\bmp}$ can not be too large. 

\section{Conclusions}
\label{sec:conclusions}
In this paper, we have investigated the relationship between the dimensionality of probe, understood as the dimension of Hilbert space, and quantum multiparameter estimation. Measurement incompatibility emerges as a critical factor in achieving the ultimate precision bounds, making strategies to overcome this challenge fundamentally important. We havw examined the potential of an enlarged Hilbert space to mitigate the quantum noise arising from this incompatibility, with a particular focus on two paradigmatic cases: the estimation problems involving two and three parameters unitarily encoded via 
$\mathfrak{su}(2)$  transformations. For these scenarios, we demonstrated that initial states exist where quantum noise can be completely eliminated without sacrificing optimal precision and scaling. We conjecture that, at least for unitary models, if the Hilbert space dimension is less than the number of parameters, simultaneous estimation becomes unfeasible. Conversely, when the Hilbert space dimension matches the number of parameters, the model exhibits the maximum AI measure. Finally, for larger Hilbert spaces, there may always be a state for which the AI measure is zero, enabling the attainment of the SLD Cramér-Rao bound.

Additionally, we evaluated the AI measure's effectiveness in assessing the 
{\em quantumness} of the quantum statistical model, understood as the difference between the Holevo-Cramér-Rao bound and the SLD one. Our findings indicate that the AI measure alone may fail to adequately quantify the gap between the two bounds, and that assessing the determinant of the QFIM is fundamental to have a precise evaluation the model’s asymptotic compatibility. Nonetheless, this quantity still plays a relevant role in such analysis, since it encapsulate the classicality of the model in one scalar quantity rather than in a matrix quantity, i.e. the Uhlmann matrix.

In a broader perspective, our insights could facilitate the design of compatible quantum statistical models by leveraging larger probes, thereby reducing the cost of achieving optimal precision. Moreover, our analysis opens potential pathways for evaluating the resilience of these approaches against classical noise.

\section*{Acknowledgements}
This work has been done under the auspices of GNFM-INdAM and has been 
partially supported by MUR through Project No. PRIN22-2022T25TR3-RISQUE. A.C. acknowledge Templeton World Charity Foundation, Inc. which supported this work through the grant TWCF62423. The opinions expressed in this publication are those of the authors and do not necessarily reflect the views of the John Templeton Foundation.

\appendix

\section{\label{app:proof1} Proof of theorem 1}

In this appendix we prove that the FIM is not invertible if the space of the outcomes is not sufficiently large. Let us consider a random variable $X$ with a finite set of outcomes of dimension $N$. We write $p_{i}(\bm\lambda)=p_{i}$ for the sake of brevity. The probability is normalized as 
\begin{equation}
\label{eq:prop1}
\sum_{i=1}^{N}p_{i} = 1.
\end{equation}
To simplify the notation in the following discussion we define
\begin{equation}
	\dd_{ij} = \frac{d}{d \lambda_{i}}p_{j} = \partial_{\lambda_{i}}p_{j} = \partial_{i}p_{j}
\end{equation}
This form a $d\times N$ matrix $\bm{\dd}$. The Fisher information matrix $\bm{\mathcal{F}}$ is defined as the $d\times d $ matrix
\begin{align*}
	\mathcal{F}_{ij}  = \sum_{k=1}^{N}  p_{k}\,\partial_{i}\!\log p_{k} \,\partial_{j}\!\log p_{k}\, p_{k} =  \sum_{k=1}^{N}  \frac{\partial_{i}p_{k}\partial_{j}p_{k}}{p_{k}}  =  \sum_{k=1}^{N}  \frac{\dd_{ik}\dd_{jk}}{p_{k}}
\end{align*}
Now we exploit \eqref{eq:prop1} and write $p_{N} = 1-\sum_{j=1}^{N-1}p_{j}$, 
which implies
\begin{equation}
	\dd_{iN} = \partial_{i} p_{N} = -\sum_{j=1}^{N-1}\partial_{i}p_{j} = -\sum_{j=1}^{N-1}\dd_{ij}
\end{equation}
In this way, the $ij$th element of the Fisher matrix can be expanded as
\begin{align*}
\mathcal{F}_{ij} & = \sum_{k=1}^{N-1}  \frac{\dd_{ik}\dd_{jk}}{p_{k}} + \frac{\Delta_{iN}\Delta_{jN}}{p_{N}} = \sum_{k=1}^{N-1}  \frac{\dd_{ik}\dd_{jk}}{p_{k}}  + \frac{\left(-\sum_{k=1}^{N-1}\dd_{ik}\right)\left(-\sum_{k=1}^{N-1}\dd_{jk}\right)}{1-\sum_{k=1}^{N-1}p_{k}} = \\
& = \sum_{k=1}^{N-1}  \left(\frac{\dd_{ik}}{p_{k}}  + \frac{\left(\sum_{m=1}^{N-1}\dd_{im}\right)}{1-\sum_{k=1}^{N-1}p_{k}}\right)\dd_{jk} 
\end{align*}
Now, if we define
\begin{equation}
\eta_{ik} = \frac{\dd_{ik}}{p_{k}}  + \frac{\left(\sum_{m=1}^{N-1}\dd_{im}\right)}{1-\sum_{k=1}^{N-1}p_{k}}
\end{equation}
we have that
\begin{equation}
\mathcal{F}_{ij} = \sum_{k=1}^{N-1}\eta_{ik}\dd_{jk} = (\bm{\eta}\cdot \tilde{\bm{\dd}})_{{ij}} 
\end{equation}
where we denoted with $\tilde{\bm{\dd}}_{ij}$the submatrix of the matrix $\bm{\dd}^T$ where we have neglected the last column. We have that $\tilde{\bm{\dd}}$ is a $N-1 \times d $ matrix defined as $\tilde{\bm{\dd}} = \dd_{kj}$, while $ \bm{\eta}$ is $d\times N-1$ matrix.  Hence we can write
\begin{equation}
\bm{\mathcal{F}} = \bm{\eta} \cdot \tilde{\bm{\dd}}
\end{equation}
To evaluate the invertibility of $\bm{\mathcal{F}}$ we have to evaluate the determinant. However, since $\bm{\eta}$ and $\tilde{\bm{\dd}}$ are not square matrices, the determinant of $\bm{\mathcal{F}}$ is not simply the product of the determinant. However, the Cauchy-Binet theorem still applies \cite{broida1989comprehensive}: if $A$ and $B$ are $n_1 \times n_2$ and $n_2 \times n_1$, then the product is a $n_1\times n_1$ matrix. Then
\begin{equation}
\det \bm{\mathcal{F}} = \sum_S \det{\bm{\eta}_S} \det{\tilde{\bm\dd}_S}
\end{equation}
where $S$ is the set of subset of $n_1$ elements in the set of $\{1,...,n_2\}$ and $A_S$ are some minors (more precise on wiki). We have that if $n_2 > n_1$, the set $S$ is empty and the determinant is $0$. Thus we have that if $ d > N-1$ the Fisher Information matrix is  not invertible. Instead, if $d \leq N-1$, then the Fisher Information matrix might be invertible. 

\section{Derivation of $\mathcal{H}_{\varphi}$ }
\label{app:threeparam}
In this appendix, we derive the $\mathcal{H}_{\varphi}$ given in Eq. \eqref{eq:Hvarphi}. Indeed, from Eq. \eqref{eq:Htimesvarphi}  it follows that
\begin{align}
H^{\times n}(\partial_{\varphi}H)  =  i \cos \theta  H^{\times n}(H^{\times}(J_{\bm{n}^{(3)}_{\theta'}}) )  = i \cos \theta  H^{\times (n+1)}(J_{\bm{n}^{(3)}_{\theta'}})
\end{align}
This means that the operator $\mathcal{H}_{\varphi}$ can be written as
\begin{align}
    \mathcal{H}_{\varphi} = -\cos(\theta) \left(\sum_{n=0}^{+\infty} f_{n-1} H^{\times (n)} (J_{\bm{n}^{(3)}_{\theta'}}) -J_{\bm{n}^{(3)}_{\theta'}}\right)
\end{align}
Since $f_{n-1} = i t^{n}/n!$, we have that the series give the exponential operator
\begin{align}
    \mathcal{H}_{\varphi} = -\cos(\theta) \left\{\exp{i t B J^{\times}_{\bm{n}^{(3)}_{\theta}}} - \mathbb{I} \right\} J_{\bm{n}^{(3)}_{\theta'}}
\end{align}
where we have used that $H^{\times}(\bullet) = B J^{\times}_{\bm{n}_{\theta}}(\bullet)$. This expression can be further simplified using the properties of the $\mathfrak{su}(2)$ algebra, i.e. 
\begin{align}
J^{\times}_{\bm{n}^{(3)}_{\theta}}(J_{\bm{n}^{(3)}_{\theta'}}) & = [J_{\bm{n}^{(3)}_{\theta}},J_{\bm{n}^{(3)}_{\theta'}}] = -iJ_{\bm{n}_{\varphi'}} \\
J^{\times}_{\bm{n}^{(3)}_{\theta}}(J_{\bm{n}_{\varphi'}}) & = [J_{\bm{n}^{(3)}_{\theta}},J_{\bm{n}_{\varphi'}}] =  i J_{\bm{n}^{(3)}_{\theta'}}
\end{align}
The algebra is closed and the recursive application of the superoperator as well. Hence we have that
\begin{equation}
J^{\times n}_{\bm{n}^{(3)}_{\theta}} (J_{\bm{n}^{(3)}_{\theta'}}) = \begin{cases}
J_{\bm{n}_{\theta'}^{(3)}} & n \text{ even} \\
-i J_{\bm{n}_{\varphi'}}  & n \text{ odd} 
\end{cases}
\end{equation}
meaning that 
\begin{align*}
	\exp\{i t B J^{\times}_{\bm{n}^{(3)}_{\theta}}\} J_{\bm{n}^{(3)}_{\theta'}} & = \sum_{n=0}^{+\infty} \frac{(i t B)^{n} J^{\times n}_{\bm{n}^{(3)}_{\theta}}(J_{\bm{n}^{(3)}_{\theta'}})}{n!} = \\
    & = \sum_{n \text{ even}}^{+\infty} \frac{(i t B)^{n} J^{\times n}_{\bm{n}^{(3)}_{\theta}}(J_{\bm{n}^{(3)}_{\theta'}})}{n!} + \sum_{n \text{ odd}}^{+\infty} \frac{(i t B)^{n} J^{\times n}_{\bm{n}^{(3)}_{\theta}}(J_{\bm{n}^{(3)}_{\theta'}})}{n!} = \\
	& = \sum_{n \text{ even}}^{+\infty} \frac{(i t B)^{n} }{n!} J_{\bm{n}^{(3)}_{\theta'}} -i \sum_{n \text{ odd}}^{+\infty} \frac{(i t B)^{n} }{n!} J_{\bm{n}_{\varphi'}}  = \\
    & = \cos Bt J_{\bm{n}^{(3)}_{\theta'}} -i \sum_{k = 0}^{+\infty} \frac{(i t B)^{2k+1} }{(2k+1)!} J_{\bm{n}_{\varphi'}}  \\
	& = \cos Bt J_{\bm{n}^{(3)}_{\theta'}} -i \sum_{k = 0}^{+\infty} \frac{i^{2k+1} (t B)^{2k+1} }{(2k+1)!} J_{\bm{n}_{\varphi'}} 
\end{align*} 
Since $i i^{2k+1} = i^{2(k+1)} = (-1)^{k+1} = (-1)^{k}(-1)$ we have that
\begin{align}
	\exp\{i t B J^{\times}_{\bm{n}^{(3)}_{\theta}}\} J_{\bm{n}^{(3)}_{\theta'}}  =  \cos Bt J_{\bm{n}^{(3)}_{\theta'}} + \sin Bt J_{\bm{n}_{\varphi'}}
\end{align}
from which we eventually Eq. \eqref{eq:Hvarphi}
\begin{align}
\mathcal{H}_{\varphi} & = -\cos\theta\left[(\cos Bt - 1)J_{\bm{n}^{(3)}_{\theta'}} +\sin Bt J_{\bm{n}_{\varphi'}}\right] = \\ 
& = 2  \sin\frac{Bt}{2}\ J_{\bm{n}^{(3)}_{2}}
\end{align}
where the vector $\bm{n}^{(3)}_{2}$ is defined in Eq. \eqref{eq:n32vector}

\section*{References}
\bibliographystyle{unsrt}
\bibliography{rdim}

\end{document}